\documentclass[runningheads]{llncs}
\usepackage[T1]{fontenc}
\usepackage{graphicx}
\usepackage{todonotes}
\usepackage{subcaption}

\newcommand{\sysname}{Open Player Socially Analytical Intelligence} 
\newcommand{\sysnameshort}{OPSAI}

\begin{document}
\title{Unlocking Open-Player-Modeling-enhanced Game-Based Learning: The \sysname{} Architecture}
\titlerunning{The Open Player Socially Analytical Intelligence Architecture}
%


\author{%
  Zhiyu Lin\inst{1} \and
  Boyd Fox\inst{2} \and
  Devon Mckee\inst{1} \and
  Sai Siddartha Maram\inst{1} \and
  Jiahong Li\inst{1} \and
  Tyler Sorensen\inst{1,3} \and
  Brian K. Smith\inst{4} \and
  Roger Azevedo\inst{5} \and
  Jichen Zhu\inst{6} \and
  Magy Seif El‑Nasr\inst{1}}

\institute{%
  University of California, Santa Cruz, Santa Cruz, CA 95064, USA\\
  \email{zlin34@ucsc.edu, dlmckee@ucsc.edu, samaram@ucsc.edu, jli906@ucsc.edu, mseifeln@ucsc.edu}\\
  \and
  Independent\\
  \email{thomas.boyd.fox@gmail.com}
  \and
  Microsoft Research, USA\\
  \email{tsorensen@microsoft.com}
  \and
  Boston College, Chestnut Hill, MA 02467, USA\\
  \email{b.smith@bc.edu}
  \and
  University of Central Florida, Orlando, FL 32816, USA\\
  \email{roger.azevedo@ucf.edu}
  \and
  ITU Copenhagen, Copenhagen, Denmark\\
  \email{jicz@itu.dk}}


%
\authorrunning{Z. Lin et al.}
%

\maketitle 
\begin{abstract}
Game-Based Learning (GBL) is a learner-engaging pedagogical methodology, yet adapting games to heterogeneous learners requires transparent, real‑time Open Player Models (OPMs).
We contribute to the community \sysname{} (\sysnameshort{}), an architecture implementing OPM beyond conceptual frameworks and validated in a GBL application.
It decouples gameplay telemetry and analysis from the game engine and automatically derives pedagogically actionable insights, supporting the transparency of computational player models while making them accessible to players.
\sysnameshort{} comprises three logical layers: a Frontend that both provides the GBL experience and collects information needed for analytics; a stateless Backend that hosts transparent analytics services producing reflective prompts, recommendations, and visualization guides; and a two-tier Log Storage that balances heavy raw gameplay data with lightweight reference indices for low-latency queries. 
By feeding analytics outputs back into the game interface, \sysnameshort{} closes the feedback loop between play and learning, empowering teachers, researchers, and learners alike. 
We further showcase \sysnameshort{} with a full deployment on the Parallel GBL environment, featuring live play traces, peer comparisons, and personalized suggestions, demonstrating a reusable blueprint for future educational games.



\end{abstract}

\section{Introduction}
Games are a powerful pedagogical medium. 
By leveraging immersive and playful mechanics, they regulate and enhance students’ intrinsic learning motivation \cite{ryan_intrinsic_2020} by simultaneously satisfying the three fundamental psychological needs enlisted in Self-Determination Theory: autonomy, competence, and relatedness \cite{andretta_effectiveness_2026}. 
Game-Based Learning (GBL) is a teaching methodology that incorporates the principles of play into educational settings with the explicit goal of fostering student engagement, motivation, and specific learning outcomes \cite{plass_foundations_2015}.
GBL offers an iterative mastery loop that reinforces competence through actionable tasks and progress tracking, supports autonomy by enabling self‑guided exploration without real‑world failure.

However, achieving these outcomes requires recognizing individual learner differences and developing strategies to analyze and adapt to them, by \textbf{modeling the players}, thereby refining both the system as a whole for individual players.
Game analytics \cite{drachen_game_2013}, which facilitate the discovery and communication of data patterns in game development and research, are indispensable for this task. 
Analysts and developers use player models to gain insights into different player types to improve gameplay \cite{drachen_game_2013},
and tailor content to individual learning needs \cite{kantharaju_tracing_2018}. 

Recent advancements in Artificial Intelligence (AI) and Machine Learning (ML), particularly data-driven methods, open new opportunities for game analysts. 
Although these models are effective at detecting patterns, they often lack explanations beyond correlation --- an essential need for stakeholders of game analytics.
Emerging research at the intersection of game analytics, eXplainable AI (XAI) \cite{tao_explainable_2023} and Interactive Machine Learning (IML) \cite{dobrovsky_approach_2016,lin_explore_2017} aims to make these techniques less opaque and more human-centric, giving rise to Open Player Modeling (OPM) \cite{zhu_open_2021}.
OPM seeks to make computational player models accessible and transparent to players themselves.

While substantial research interest has developed within OPM, including 
Affective Computing \cite{yannakakis_affective_2023},
e-sports \cite{margetis_visual_2023},
and GBL \cite{maram__2024},
most work focuses on using OPM as a conceptual framework.
There lacks a clear architecture that discusses how such models and other supporting components are developed within the context of games that combine complex data analytics techniques with an infrastructure for instrumentation collection and modeling toward better learning.



In this work, we establish the \sysname{}  (\sysnameshort{}) architecture, a novel OPM-compliant architecture that seamlessly integrates real‑time telemetry with player modeling within the context of GBL, 
thereby enabling dynamic, learner‑centric feedback loops and visualizations empowered with modular and scalable game analytics pipelines.

Our contributions are as follows:
\begin{itemize}
    \item \sysnameshort{} architecture: an OPM-compliant, end‑to‑end pipeline that decouples gameplay telemetry from the game engine, enriches it with contextual metadata, and automatically derives pedagogically actionable learning metrics in real time, thereby benefiting researchers, developers, and learners alike and empowering both the GBL experience and the analytics.
    \item Exemplar deployment on the Parallel GBL experience: a complete implementation that demonstrates live visualizations, learner-centric feedback loops, and modular, pedagogically focused analytics for educational research.
\end{itemize}

\section{Design of \sysnameshort{}}
OPM requires an analytics framework that not only capture and process gameplay telemetry but also translate it into pedagogical insights for both researchers and learners, a capability that conventional analytics platforms, which stakeholders involves mainly developers and investors instead of users, lack out-of-the-box.
\sysnameshort{} addresses this gap with an end-to-end pipeline that ingests raw gameplay data in real time, optionally enriches it with contextual metadata and sensory inputs, and automatically derives key learning metrics in its analytics components, thereby enabling rapid, data-driven insights without adding load to the game engine or the player.
The architecture further provides structured support for modular support of 
toolkits that feeds these derived metrics back into the game, empowering live visualizations and actionable guidance that inform teachers and researchers with timely, socially context‑rich insights. 
This is achieved by defining them as stateless components and decoupling analytics from the core game engine, allowing horizontal scaling, optimized data storing and retrieval, and extension to other GBL domains without extensive redesign.

To achieve this, the \sysnameshort{} establishes three logical layers:
\begin{itemize}
    \item the \textbf{Frontend}: the stakeholder-facing component that delivers the GBL experience and displays visualizations and insights.
    \item the \textbf{Backend}: the analytics engine that maps data to actionable insights and reflections.
    \item the \textbf{Log Storage}: the persistent storage layer responsible for efficient data storage, indexing, and processing.
\end{itemize}
Each logical layer encapsulates a distinct responsibility, while well‑defined APIs and data structures governs communication across \sysnameshort{}.

\subsection{Frontend}
The Frontend presents both the game and the analytics produced by \sysnameshort{}.
At its core is a generic data model that exposes the current game state, player actions, and derived metrics in a game-agnostic format, acting as the \textit{Store} for both the game and the analytics engine.
A stateful client renders the game using the Store as its source of context and exposes controls that manipulate the game state in accordance with the Game Logic engine.
The Store is then serialized and streamed to the Backend in real time, ensuring that gameplay events are captured immediately.
After analysis, the Frontend hosts additional modules that expose results to learners as needed --- for example, a visualization module that consumes analytic payloads from the Backend and displays actionable insights such as play traces, personalized recommendations, and peer‑comparison dashboards, thereby closing the interactive loop from gameplay to immediate feedback.

\subsection{Backend}
The Backend orchestrates the analytics engine of \sysnameshort{}, and mediates the other logical layers of the architecture.
It receives real-time event streams from the Frontend, mediates persistence in Log Storage, and runs the analytics services described by OPM to generate insights such as reflective prompts, personalized learning recommendations, and visualization guides.
Structured log data distilled from gameplay interactions is used to compute player-specific metrics (for example, movement patterns, decisions of players, board-state trajectories), and the results are returned to the Frontend as an analytics payload.
Unlike the stateful Frontend, the Backend exposes a stateless, API-driven interface that allows on-demand analytics requests.
All logic is isolated from the game engine, enabling horizontal scaling of additional analytics components and rapid iteration without affecting GBL experience.

\subsection{Log Storage}
The Log Storage layer efficiently stores and presents data to the other layers.
Because real-time play traces generate large volumes of data, the architecture adopts a hybrid persistence layer that separates lightweight metadata from bulky raw event data while providing efficient query paths for analytics.
A log database stores the full representation of each session, including timestamped events, board- states, user decisions, and supports enrichment of records using optional data such as eye-tracking and external feedback.
A Log Reference table enables rapid retrieval of specific logs or queries that match particular criteria without scanning the entire database; this is achieved by a preprocessing engine that analyses logs and creates lightweight reference entries, while also enriching log artifacts for the log database.
The design delivers low-latency lookups for the analytics engine while allowing log size and fields to scale without incurring high read costs, thereby supporting a robust, scalable gameplay and analytics experience for learners.

\subsection{Exemplar Integration}

\begin{figure}
    \centering
    \includegraphics[width=0.85\linewidth]{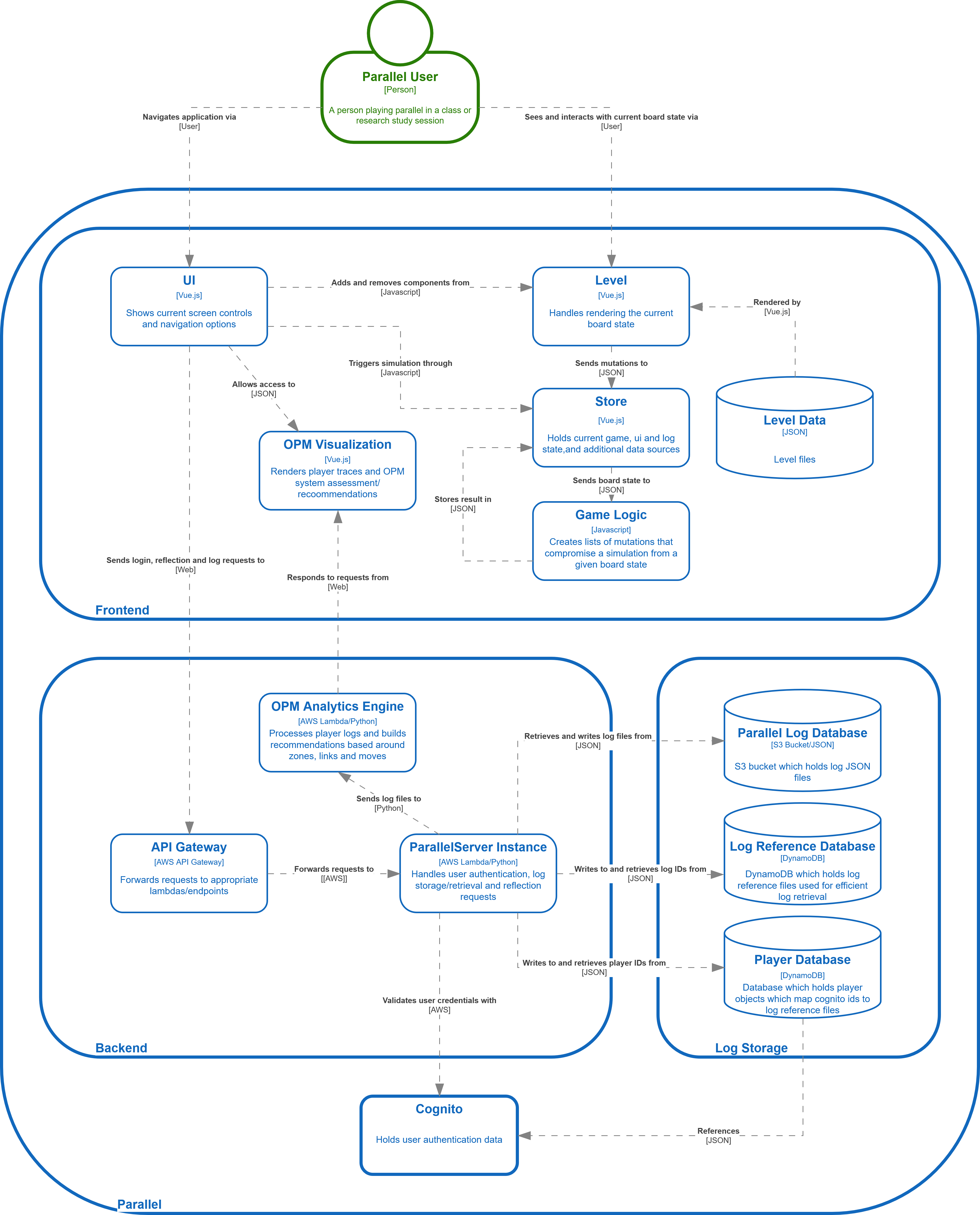}
    \caption{An instantiation of \sysnameshort{} in \textit{Parallel}.}
    \label{fig:arch}
\end{figure}

We now showcase our architecture in detail with an exemplar integration with \textit{Parallel} the GBL experience.

\begin{figure}
    \centering
    \includegraphics[width=0.6\linewidth]{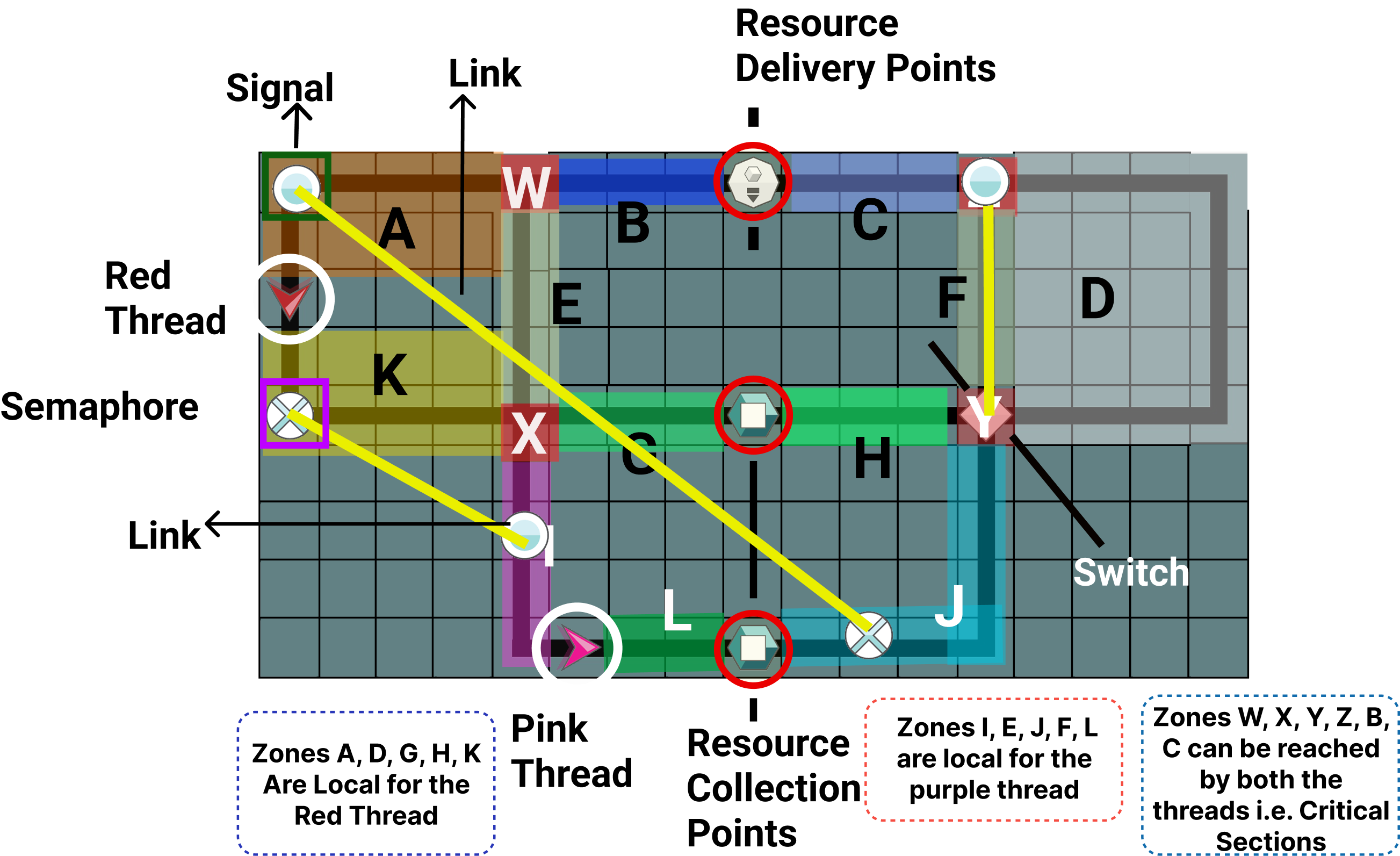}
    \caption{Game elements in \textit{Parallel}. See \cite{valls-vargas_graph_2017} for further details.}
    \label{fig:parallel_game}
\end{figure}

Parallel \footnote{\url{playparallel.com}} is a 2D spatial puzzle game originally developed by Valls-Vargas et al. \cite{valls-vargas_graph_2017}, aimed to help students learn parallel programming.
As illustrated in Figure \ref{fig:parallel_game}, students are required to place semaphores, signals and connect the signals with semaphores to coordinate the movement of arrows representing threads. 
To solve levels in this game, students need to demonstrate understanding of concepts like mutual exclusion, critical sections and how semaphores and signals work in parallel programming. 

A component diagram of our instantiation of \sysnameshort{} within Parallel is provided in figure \ref{fig:arch}.

\begin{figure}
    \centering
    \includegraphics[width=\linewidth]{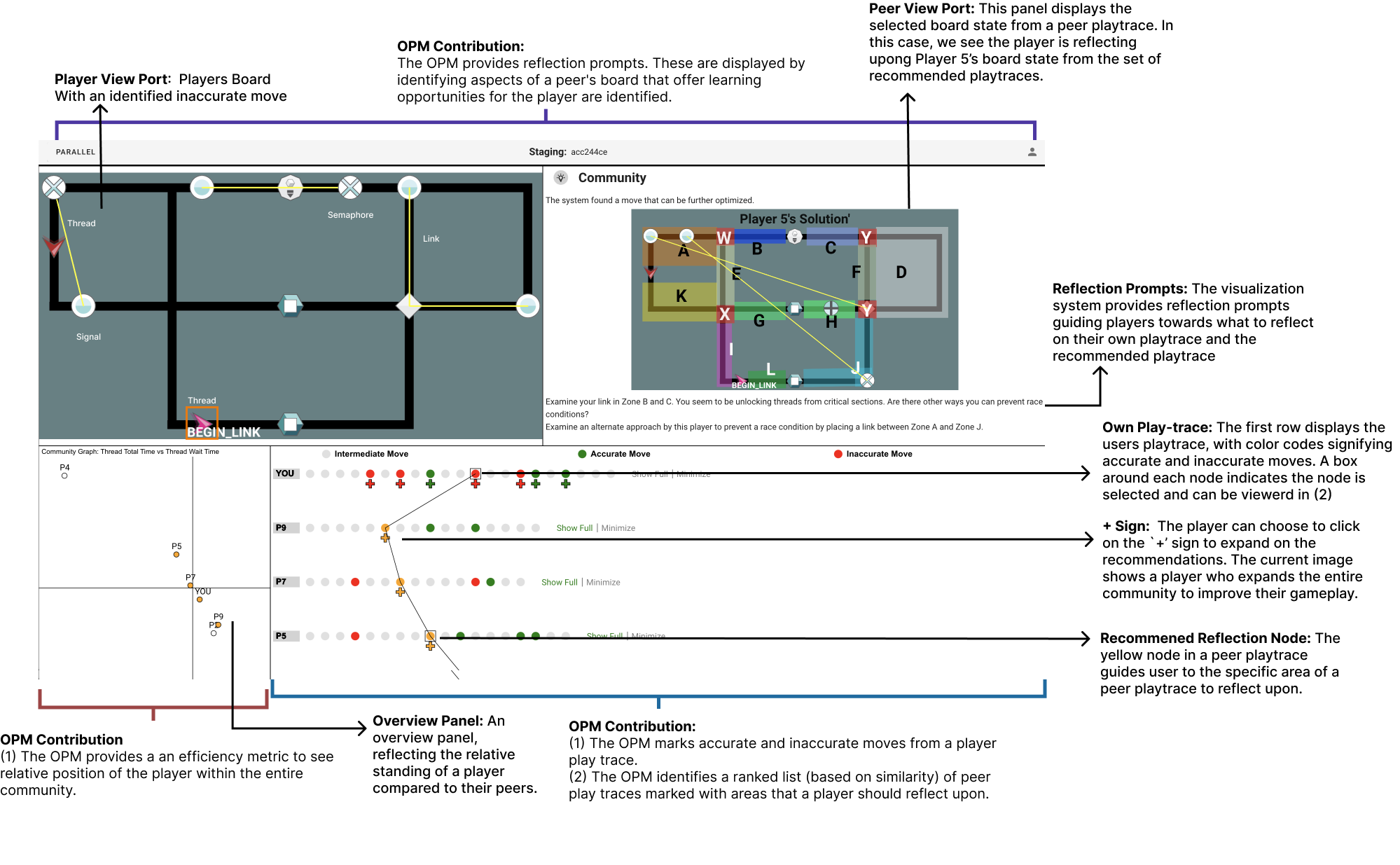}
    \caption{Exemplar visualization interface.}
    \label{fig:parallel_viz}
\end{figure}


\paragraph{Implementation of Frontend.}
A web‑based game application built with Vue.js serves as the game engine for the Parallel experience.
In this instance, the universal game data structure i.e. the \textit{Store} is implemented as JSON files, holding both the game levels and current gameplay states.
The visualization component, also implemented in Vue.js, renders actionable insights upon request.
Thanks to \sysnameshort{}, efficient analysis of not only the players’ experience but also other players’ enables display of insights based on social elements (See figure \ref{fig:parallel_viz}, such as play traces from the player’s most recent games versus peers with similar actions, using analytic information from the Backend. These insights are directly available within the game interface, closing the interactive loop from gameplay to immediate feedback.

\paragraph{Implementation of Backend Server.}
Our architecture enabled efficient integration of the analytics engine described in Zhu et al. ’s work \cite{zhu_open_2021}, supplying reflective prompts and recommendations to the Frontend.
It consumes structured log data from the Log Storage layer and generates visualization guides and personalized learning recommendations.

\paragraph{Implementation of Log Storage.}
Three solutions utilizing different software stacks are used to implement the architecture, deployed on Amazon Web Services (AWS). 
The Log Database is deployed as an object‑oriented database utilizing Amazon S3, which stores the full JSON representation of each session in widely different data structures. 
The Log Reference Database, implemented with AWS DynamoDB, holds lightweight entries that map session identifiers to S3 object keys, enabling rapid retrieval of a specific user’s log or logs that match particular criteria without the costly operation of scanning the entire S3 bucket. 
To enable this hybrid design, an AWS Lambda is employed; a serverless Python routine mediates this process, fulfilling requirements enlisted in the high-level architecture.


By structuring the system into these three decoupled logical layers, \sysnameshort{} achieves a clean separation of responsibility: the front end handles user interaction and on-the-fly simulations; the Backend provides stateless analytics logic and mediate other components; the storage layer offers scalable persistence. 
This also enables external components to easily plug into each of the three layer, without need of substantially changing the implementation of any other components of the instantiation.  
In summary, \sysnameshort{} not only supports the educational and analytical goals of the GBL experience but also enables extensibility for future analytics and research goals into other GBL domains.

\section{Discussion}


Our framework closes the loop from game telemetry to actionable feedback, and its modular design allows additional functionalities to be integrated seamlessly, making it well suited for the current evolving landscape of AI in education.

We see two promising directions: (1) Adding a chatbot AI agent that connects the visualizer to learners through a dialogue interface, supporting multi‑round interactions and active learning; 
(2) Integrating additional pedagogical tools and a dynamic multi-agent system that learns and adapts to new capabilities on-the-fly within the framework, thereby leveraging its horizontal scalability.
\section{Conclusion}
In this work we introduced the \sysname{} architecture (\sysnameshort{}), an Open Player Modeling-compliant end-to-end analytics pipeline that closes the loop among players, educators, and researchers by bridging raw telemetry and actionable pedagogical feedback.
Our exemplar integration in the Parallel Game-Based Learning experience demonstrates the ability of the architecture to support delivery of live visualizations and recommendation overlays, while its modular design permits well decoupling of responsibilities and effortless extension to other GBL contexts.
By validating our architecture in a real GBL setting, we show \sysnameshort{}'s potential as a blueprint that bridges the gap between concepts and the nuanced requirements of implementing such game analytics systems.
\bibliographystyle{splncs04}
\bibliography{references}
%




\end{document}